\documentstyle[prl,aps,psfig,multicol]{revtex}
\begin{document}
\title{``Forbidden'' Transitions Between Quantum Hall and Insulating Phases\\ in p-SiGe Heterostructures}
\author{M.~R. Sakr, Maryam Rahimi, and S.~V. Kravchenko}
\address{Physics Department, Northeastern University, Boston, Massachusetts 02115}
\author{P.~T. Coleridge, R.~L. Williams, and J. Lapointe}
\address{Institute for Microstructural Sciences, National Research Council of Canada, Ottawa, Ontario, K1A OR6 Canada}
\date{\today}
\maketitle
\begin{abstract}
We show that in dilute metallic p-SiGe heterostructures, magnetic field can cause multiple quantum~Hall-insulator-quantum~Hall transitions.  The insulating states are observed between quantum Hall states with filling factors $\nu=1$ and 2 and, for the first time, between $\nu=2$ and 3 and between $\nu=4$ and 6.  The latter are in contradiction with the original global phase diagram for the quantum Hall effect.  We suggest that the application of a (perpendicular) magnetic field induces insulating behaviour in metallic p-SiGe heterostructures in the same way as in Si MOSFETs.  This insulator is then in competition with, and interrupted by, integer quantum Hall states leading to the multiple re-entrant transitions.  The phase diagram which accounts for these transition is similar to that previously obtained in Si MOSFETs thus confirming its universal character.
\end{abstract}
\begin{multicols}{2}
In many dilute two-dimensional (2D) electron and hole systems, reentrant transitions between quantum Hall (QH) and insulating states are often observed in a perpendicular magnetic field.  In silicon metal-oxide-semiconductor field-effect transistors (MOSFETs) and p-SiGe heterostructures, insulating states appear between neighboring integer QH states (see, {\it e.g.}, Refs.[1-10]), while in n- and p-GaAs/AlGaAs heterostructures, insulating states are seen between fractional QH states ({\it e.g.}, \cite{goldman90,jiang91}).  The origin of these reentrant transitions is not well understood.  In some publications, they have been attributed to field-induced formation/melting of a Wigner crystal \cite{kravchenko91,goldman90}; others \cite{shashkin93} have tried to explain them on a basis of the global phase diagram for the quantum Hall effect suggested by Kivelson, Lee, and Zhang \cite{kivelson92}.  The latter, however, allows for direct transitions to the insulator only from the integer QH state with the Hall conductivity $\sigma_{xy}=1\,e^2/h$ (thereafter referred to as the QH state ``1''; see Fig.~1~(a)) or from fractional QH states with Hall conductivities $\sigma_{xy}=1/3$, $1/5$, $1/7\,e^2/h$..., while experimentally, direct transitions to the insulator from higher integer QH states were observed in some 2D systems \cite{diorio90,pudalov93,shashkin93,kravchenko95,hilke00}, as well as ``forbidden'' transitions between insulator and fractional QH states ``2/5'', ``2/7'', and ``2/9'' \cite{jiang91}.  In Ref.\cite{kravchenko95}, a phase diagram for integer QH effect was constructed based on experimental data in Si MOSFETs, and it was shown that, being generalized to the case of the {\em fractional} QH effect, this diagram is consistent with the above-mentioned ``forbidden'' transitions between insulator and fractional QH states.

In p-SiGe heterostructures an insulating phase frequently appears between the QH states ``1'' and ``2'' \cite{fang92,dorozhkin95,dunford96,coleridge97a}.  The scaling behaviour at the transition between this insulating state and neighboring QH state ``1'' matches that for the transition between the ``last'' ($\sigma_{xy}=1\,e^2/h$) QH state and the high field ``$\sigma_{xy}=0$'' Hall insulating state \cite{coleridge00,possanzini01} so one possibility is that this is a premature transition into the Hall insulating state and, as such, does not violate the global phase diagram.  Furthermore, when the global phase diagram is modified to accommodate the spin degree of freedom \cite{fogler95} as shown schematically in Fig.~1~(b), direct transitions to the insulator are allowed from both ``1'' and ``2'' QH states.

Here we present data obtained in a dilute p-SiGe heterostructure, where not only is there an insulating phase between QH states at Landau filling factors $\nu=1$ and 2, but also between $\nu=2$ and 3 and between $\nu=4$ and 6. Here the filling factor $\nu$ is defined, as usual, as $p_sh/eB_\perp$ where $p_s$ is the hole density and $B_\perp$ is the component of the magnetic field perpendicular to the 2D plane. The spin degeneracy is two and there is no valley degeneracy so insulating phases appearing with $\nu>2$ represent a clear violation of the genuine global phase diagram \cite{kivelson92} in this 2D system.  A modified phase diagram, which would incorporate these ``forbidden'' field-induced transitions, is shown schematically Fig.~1~(d).  Its topology is similar to that of the diagrams determined experimentally by Hilke {\it et al.} \cite{hilke00} in p-Ge/SiGe quantum wells and numerically within a tight-binding model by Sheng and Weng \cite{sheng00}, see Fig.~1~(c).  However, the essential peculiarity of our phase diagram is the non-monotonic (``bumpy'') QH-insulator phase boundary, with maxima at integer filling factors and minima at semi-integer ones (similar to those in Ref.\cite{kravchenko95} for Si MOSFETs).  This shape of the phase diagram allows for the multiple field-induced QH-insulator-QH transitions observed in our experiment.

In p-SiGe the holes reside in a 40~nm 13\% Si(Ge) quantum well, strained because it is lattice matched to 
\vbox{
\vspace{-3.5mm}
\hbox{
\hspace{0.10in}
\psfig{file=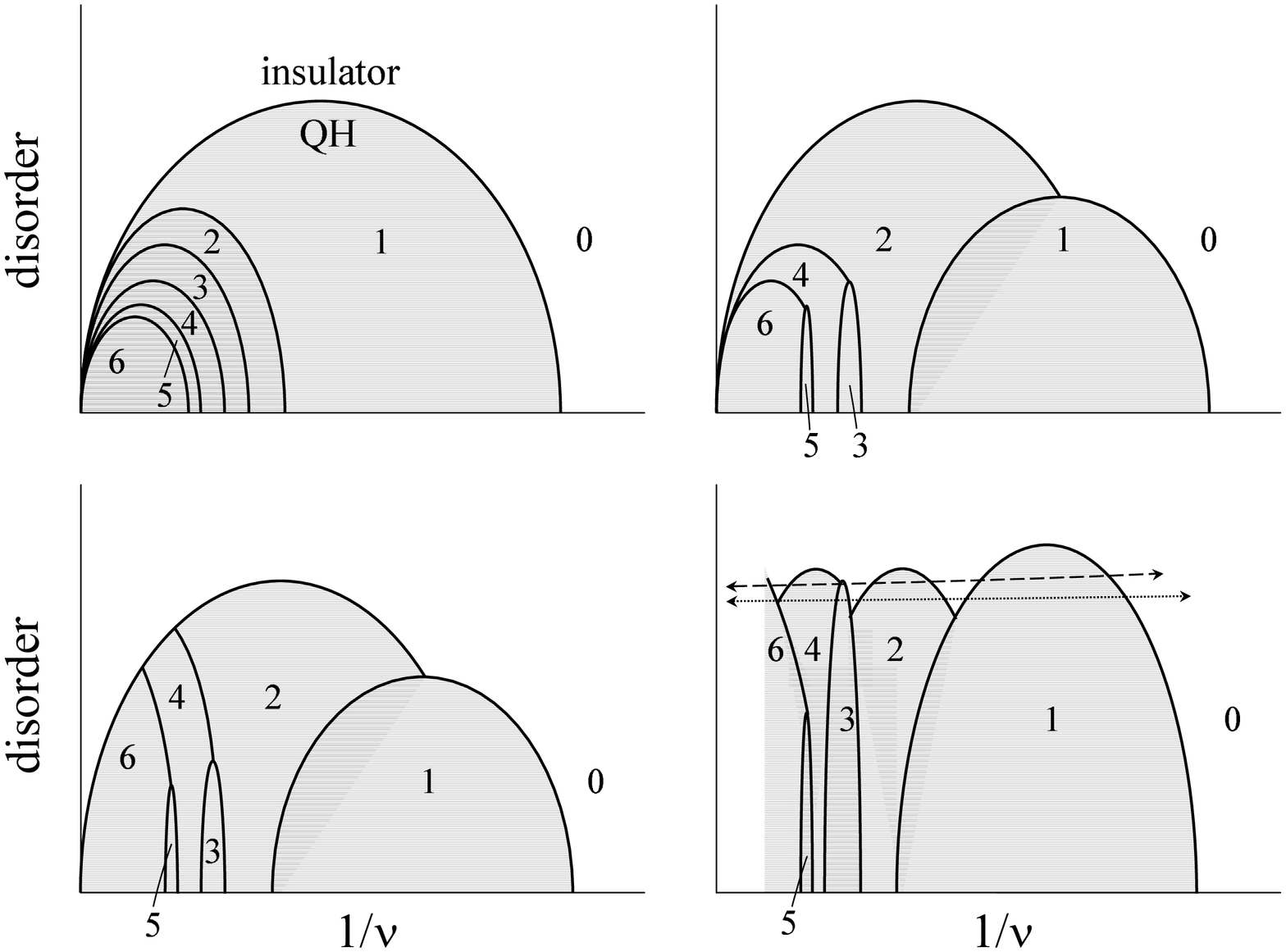,width=2.8in,bbllx=.5in,bblly=1.25in,bburx=7.25in,bbury=9.5in,angle=-90}
}
\vspace{0.4in}
\hbox{
\hspace{-0.15in}
\refstepcounter{figure}
\parbox[b]{3.4in}{\baselineskip=12pt \egtrm FIG.~\thefigure.
Schematic phase diagrams for the integer quantum Hall effect.  Shaded regions are quantum Hall states with zero longitudinal conductivity, $\sigma_{xx}$, and non-zero Hall conductivity, $\sigma_{xy}$, values of which (in units of $e^2/h$) are indicated by numbers.  White regions are insulating states with $\sigma_{xy}=\sigma_{xx}=0$.  Lines separating quantum Hall and insulating regions are extended states with nonzero $\sigma_{xx}$ and $\sigma_{xy}$.  (a)~Global phase diagram for spinless electrons suggested in Ref.~\protect\cite{kivelson92}.  (b)~Phase diagram modified for electrons with spins \protect\cite{fogler95}.  (c)~Numerical phase diagram for electrons with spins \protect\cite{sheng00}.  The upper extended states merge with the lowest one allowing for direct transitions between upper QH states (``6'' and ``4'' in this figure) and insulator.  (d)~Schematic phase diagram, similar to that obtained in Si MOSFETs in Ref.\protect\cite{kravchenko95}, consistent with our experimental results.  We do not show the low-magnetic-field part of the diagram since the behaviour of the extended states at low fields is beyond the scope of this paper.  Dashed and dotted lines show field-induced transitions between QH and insulating states corresponding to the $\rho_{xx}(B)$ dependences of Fig.~2~(a) and (b) (``6''-insulator-``4''-``3''-insulator-``2''-insulator-``1''-insulator and ``3''-insulator-``2''-insulator-``1''-insulator) respectively.
\vspace{0.10in}
}
}
}
Si. Strain removes the heavy hole/light hole degeneracy so the holes experience a strong spin-orbit coupling and have almost pure $\mid M_J \mid =\frac{3}{2}$ symmetry.  Although the spin degree of freedom is quenched the splitting of the $M_J=\pm\frac{3}{2}$ levels can be well described by a Zeeman splitting that is large and anisotropic \cite{fang92}. To a first approximation it depends, like the orbital cyclotron energy, only on the perpendicular component of the magnetic field. 

The results presented were obtained from one sample but other samples taken from the same wafer \cite{sampledetails} (and measured in a different experimental set-up) gave essentially identical data. The hole density of approximately 1.1$\times10^{11}$~cm$^{-2}$  can be compared with the critical densities for the $B=0$ metal-insulator transition, observed in very similar samples \cite{coleridge97b}, between 0.8 and 1.0$\times$10$^{11}$~cm$^{-2}$. The ratio $r_s\equiv1/(\pi p_s)^{1/2}a_B$ is about 5 (here $a_B$ is the effective Bohr radius in the semiconductor) so Coulomb interaction effects should be large.

\vbox{
\vspace{5mm}
\hbox{
\hspace{0.4in}
\psfig{file=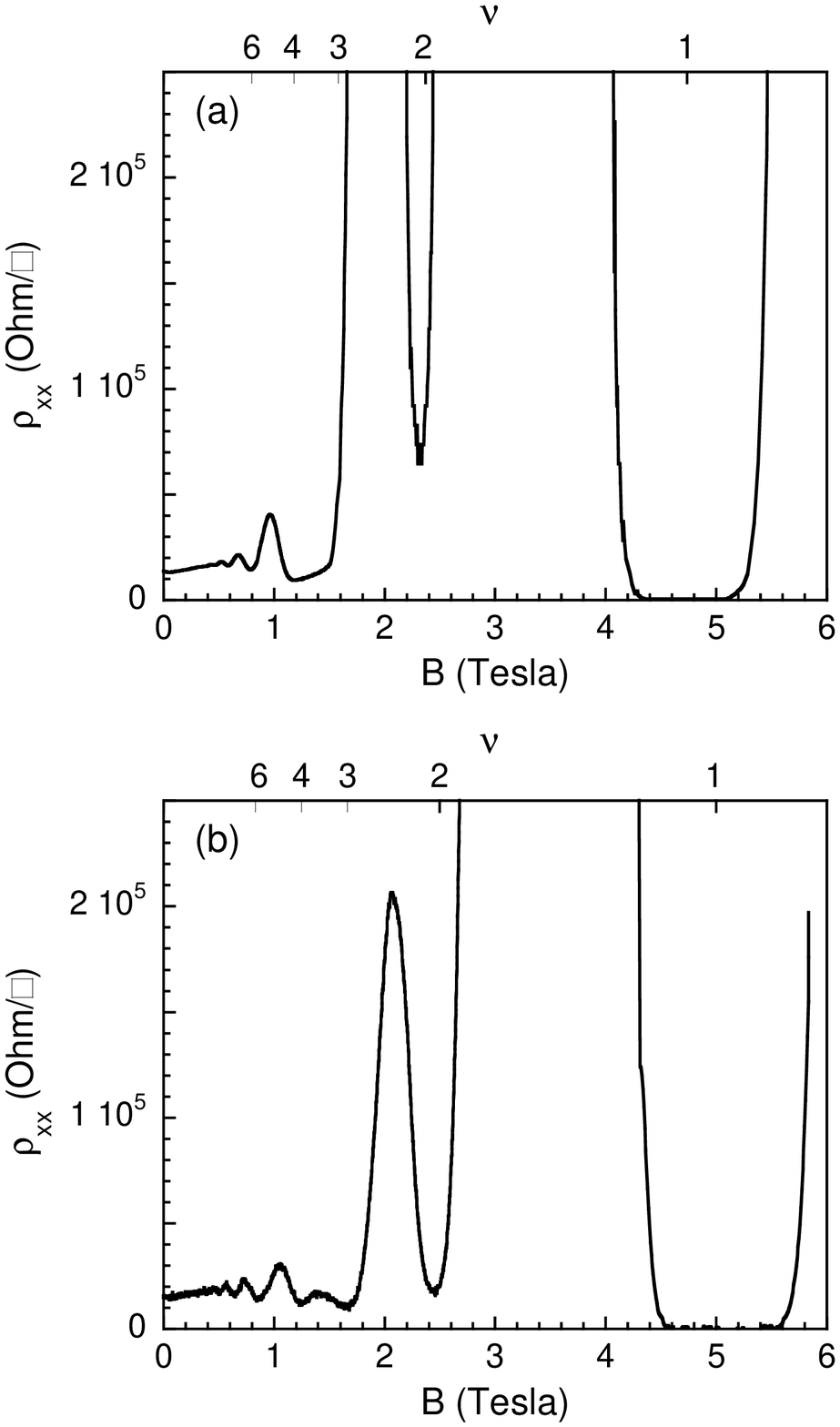,width=2.3in,bbllx=.5in,bblly=1.25in,bburx=7.25in,bbury=9.5in,angle=0}
}
\vspace{0.3in}
\hbox{
\hspace{-0.15in}
\refstepcounter{figure}
\parbox[b]{3.4in}{\baselineskip=12pt \egtrm FIG.~\thefigure.
Resistivity as a function of perpendicular magnetic field at $T=35$~mK.  (a)~$p_s=1.14\times10^{11}$~cm$^{-2}$; (b)~$p_s=1.24\times10^{11}$~cm$^{-2}$.
\vspace{0.10in}
}
}
}
\vbox{
\vspace{3mm}
\hbox{
\hspace{10.3mm}
\psfig{file=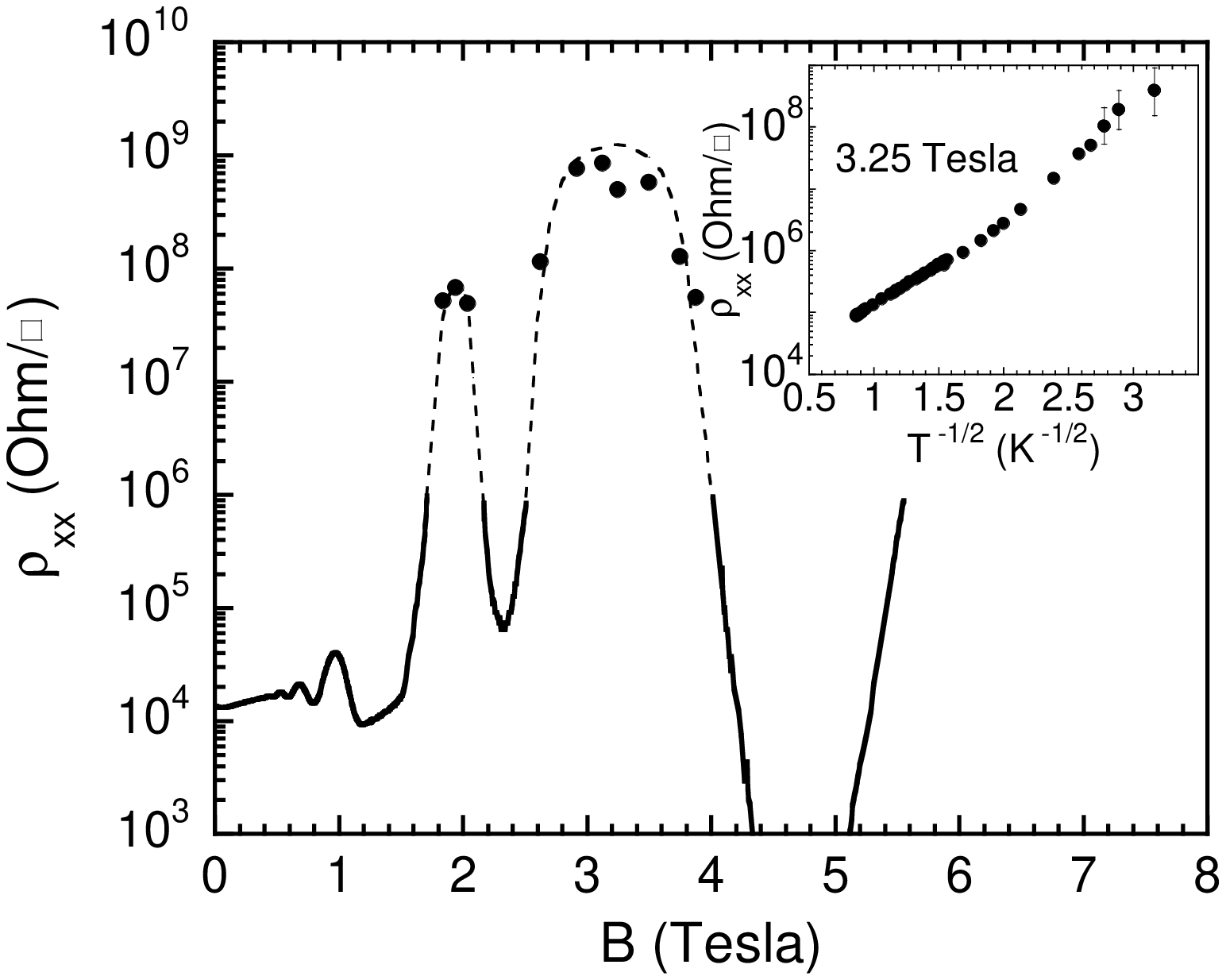,width=2.0in,bbllx=.5in,bblly=1.25in,bburx=7.25in,bbury=9.5in,angle=0}
}
\vspace{-25mm}
\hbox{
\hspace{-0.15in}
\refstepcounter{figure}
\parbox[b]{3.4in}{\baselineskip=12pt \egtrm FIG.~\thefigure.
Same as Fig.~2~(a) with resistivity plotted on a log scale.  The inset shows resistivity of the peak at $B=3.25$~tesla {\it vs} $T^{-1/2}$.
}
}
}

Measurements were made in a rotator-equipped dilution refrigerator with temperatures down to approximately 30~mK.  The Hall bar sample had a width of 200~$\mu$m and the AC measuring currents were kept small (less than 100 pA at the lowest temperatures) to avoid heating.  In the case of very high resistances, we switched to a DC technique slowly sweeping current, $I$, between $-15$~pA and $+15$~pA and measuring the voltage, $V$, for every value of the magnetic field.  The resistance was then determined as the derivative $dV/dI$ at $I=0$.

Figure~2~(a) shows the longitudinal resistivity $\rho_{xx}$ as a function of magnetic field at a temperature of 35~mK (the same dependence is shown in Fig.~3 on a logarithmic scale).  There is a well defined $\nu=1$ quantum Hall minimum of the resistance at 4.7~tesla, corresponding to a density of $1.14\times10^{11}$~cm$^{-2}$, which agrees well with the periodicity of the low field Shubnikov-de~Haas oscillations.

The two large peaks in the resistivity, at magnetic fields of about 2 and 3.2~tesla ($\nu\approx\frac{5}{2}$ and $\frac{3}{2}$ respectively)
\vbox{
\vspace{-14.2mm}
\hbox{
\hspace{10.3mm}
\psfig{file=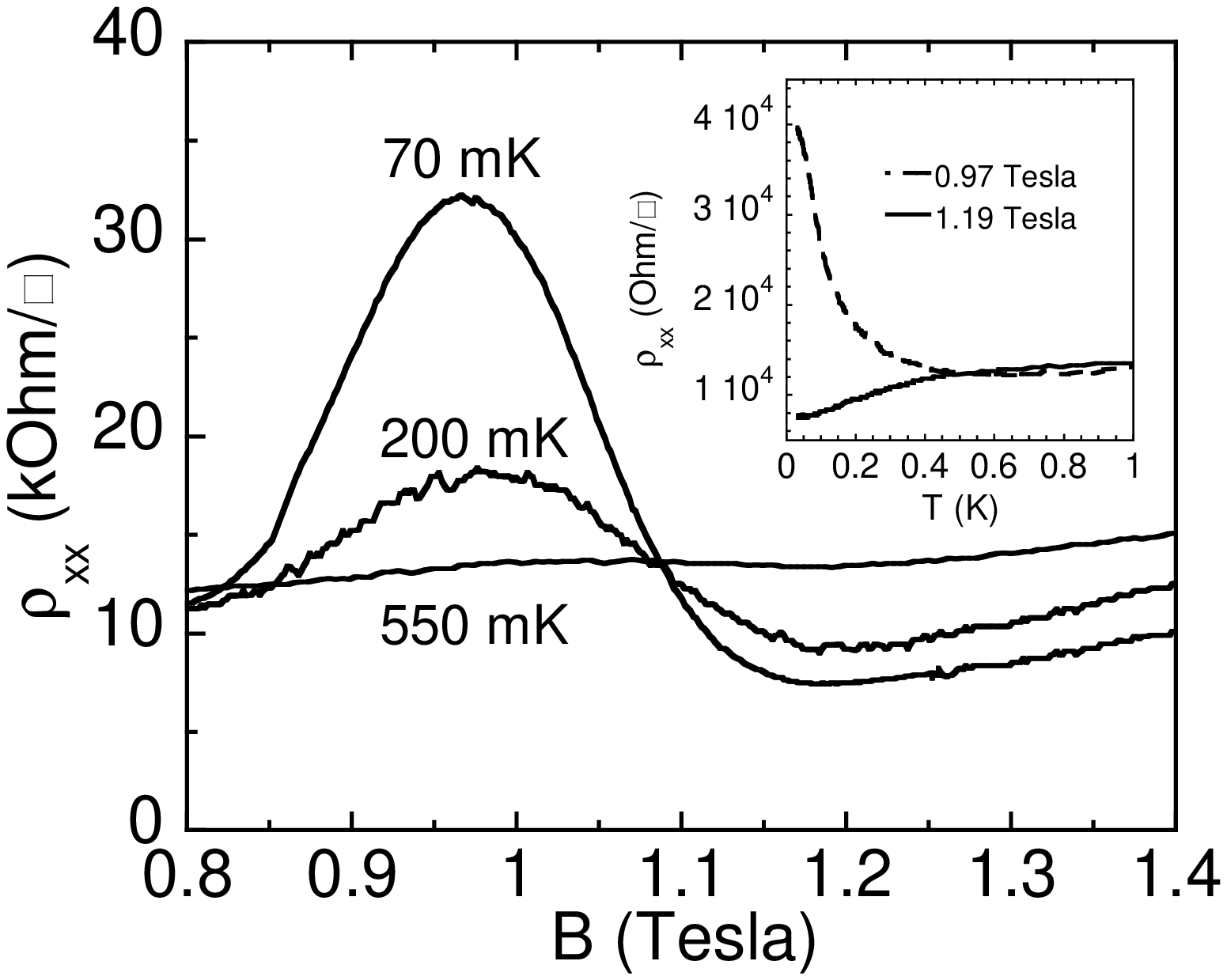,width=2.0in,bbllx=.5in,bblly=1.25in,bburx=7.25in,bbury=9.5in,angle=0}
}
\vspace{-7mm}
\hbox{
\hspace{-0.15in}
\refstepcounter{figure}
\parbox[b]{3.4in}{\baselineskip=12pt \egtrm FIG.~\thefigure.
Resistivity as a function of perpendicular magnetic field at three temperatures.  The inset shows temperature dependences of $\rho_{xx}$ at $B=0.97$~tesla (peak) and $B=1.19$~tesla (minimum).
\vspace{0.1in}
}
}
}
both show an exponentially strong increase of resistance as the temperature decreases (see {\it e.g.} the inset to Fig.~3), indicative of insulating behaviour.  In addition (see Fig.~4) the peak around 1~tesla, {\it i.e.} between $\nu=4$ and 6, also has an abnormal ``insulating'' temperature dependence.  For ordinary Shubnikov-de~Haas oscillations, the temperature dependence of the oscillation amplitude is determined by the thermal damping factor, $X_T/\sinh(X_T)$, where $X_T=2\pi^2k_BT/\hbar\omega_c$. The thermal damping factor is approximately 0.92 at 200~mK, close to the saturation value of one. As the temperature is further decreased, however, it is found experimentally that 
the height of this peak and the depth of the adjacent QH minimum at 1.19~tesla both continue to change substantially (see in the inset to Fig.~4).  The peak in particular increases by a factor of about five.  This is characteristic of insulating behaviour and we therefore consider the $\rho_{xx}(B)$ dependence shown in Fig.~4 as indicative of another reentrant transition between an insulator and a QH state, this time at $\nu\approx4$.

The $\rho_{xx}(B)$ dependence is also shown in Fig.~2~(b) for the same sample after illumination was used to increase the hole density by about 9~\%.  The two insulating peaks at approximately 2 and 3.5~tesla have decreased in amplitude and the third peak at about 1~tesla now displays a more ``usual'' (non-insulating) behaviour.  A small increase in the hole density therefore both reduces the number of insulating peaks observed and weakens the insulating character of the remaining peaks.

In other p-SiGe samples where a $\nu=\frac{3}{2}$ insulating phases is observed \cite{dorozhkin95,coleridge97a} tilting the magnetic field produces a modest enhancement of the insulating character.  A similar effect is observed here: the amplitude of all three insulating peaks is increased by tilting the magnetic field (an example is shown in Fig.~5)

Evidence has been presented showing three (two for the illuminated sample) separate re-entrant transitions from QH to insulating phases.  These all have the same general behaviour as the $\nu=\frac{3}{2}$ insulating phase seen in other p-SiGe samples. There is, at present, no generally accepted explanation for the origin of this effect.  It can, however, be compared with corresponding behaviour in 
\vbox{
\vspace{-14.5mm}
\hbox{
\hspace{10.3mm}
\psfig{file=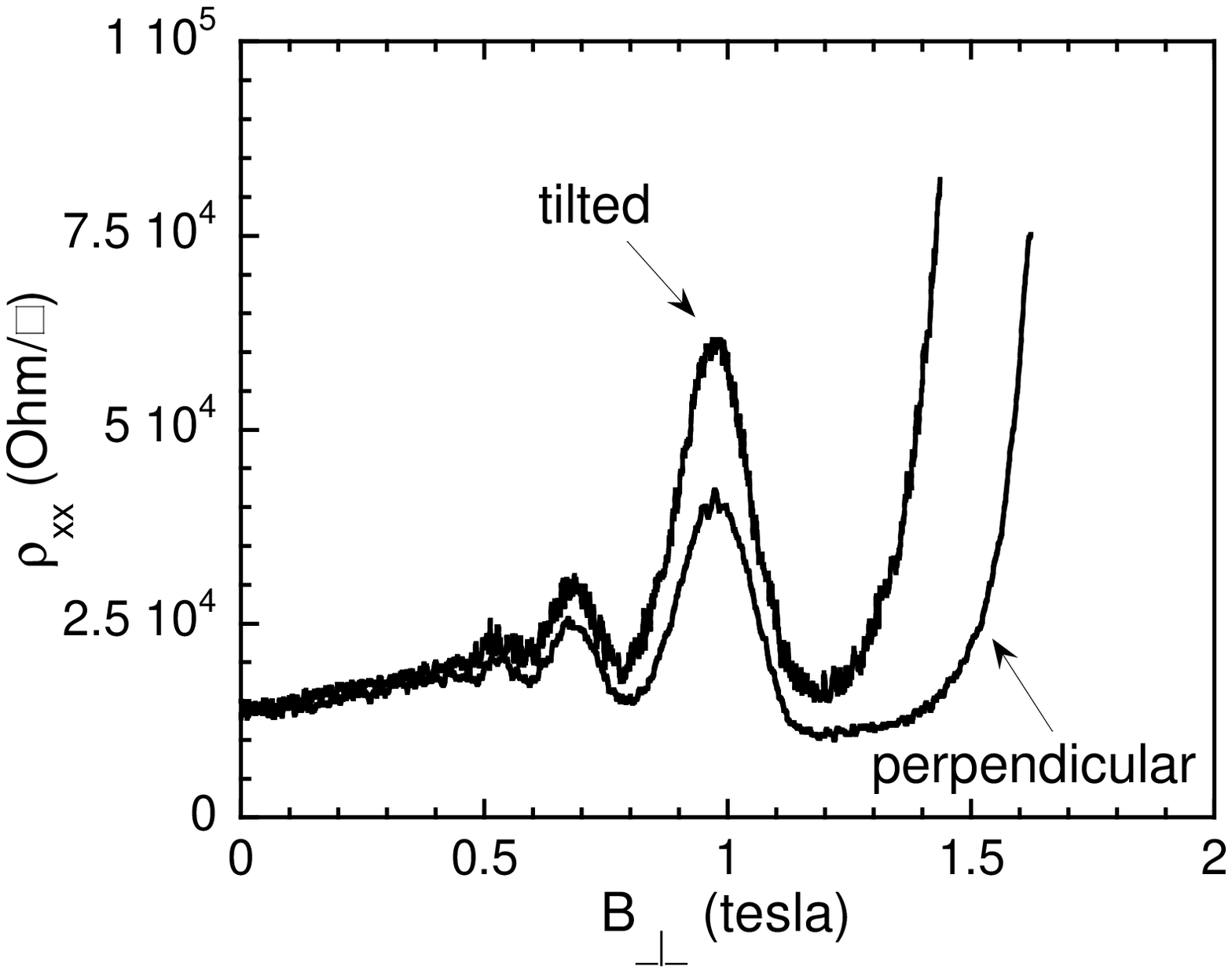,width=2.0in,bbllx=.5in,bblly=1.25in,bburx=7.25in,bbury=9.5in,angle=0}
}
\vspace{-7mm}
\hbox{
\hspace{-0.15in}
\refstepcounter{figure}
\parbox[b]{3.4in}{\baselineskip=12pt \egtrm FIG.~\thefigure.
$\rho_{xx}$ {\it vs} $B_\perp$ for magnetic field directed perpendicular to the 2D plane and tilted by 75$^o$, as labeled.  $T=70$~mK.
\vspace{0.1in}
}
}
}
dilute Si MOSFETs [1-5].  For Si MOSFETs which are on the metallic side of the $B=0$ metal-insulator transition it is known \cite{abrahams01} that magnetic fields of the order of a few tesla (either perpendicular, parallel, or tilted) induce insulating behaviour by aligning the electrons' spins and quenching the spin degree of freedom.  In p-SiGe samples, the strong spin-orbit coupling means parallel magnetic fields have relatively little effect but it can be assumed that {\it perpendicular} magnetic fields will act in the same way and induce insulating behaviour. This insulator is then in competition with, and may be interrupted by, integer QH states at yet higher magnetic fields leading to the multiple re-entrant QH-insulator-QH transitions that are seen experimentally.  An intriguing question remains about the nature of a very similar re-entrant insulating state observed around {\em fractional} filling factors in dilute GaAs/AlGaAs heterostructures \cite{jiang91}.  We note once again that the field-induced QH-insulator transitions around fractional filling factors in these systems are consistent with the experimental phase diagram for the integer QH effect \cite{kravchenko95}, generalized to the case of the fractional QH effect.

In Si MOSFETs the fact that the onset of insulating behaviour depends on the {\em total} magnetic field demonstrates very directly that the spins play a crucial role. The absence, in p-SiGe, of a similar response to parallel fields confirms this point of view. The spin degree of freedom is already suppressed by the spin orbit coupling and cannot be further reduced. The small residual response is a second order effect. It could be associated with a minor realignment of the Landau levels or alternatively changes in wavefunction overlap produced by the diamagnetic energy shift associated with parallel fields.

It should be noted that for all the data presented here the Landau level broadening is large.  At $B=0$, $k_Fl$ is of order two (here $k_F$ is the Fermi wave number and $l$ is the mean free path) and it is known that the transport and quantum lifetimes are approximately equal in this material \cite{coleridge96} so the Landau level broadening is of order half the Fermi energy.  For the higher order insulating phases ({\it i.e.} those observed at approximately 1 and 2~tesla), the Landau level broadening is therefore of the order of the cyclotron spacing and the insulating phases appear not out of well defined quantum Hall states, but rather out of states where disorder induces a strong Landau level mixing. This suggests that except at the highest magnetic fields, it is more appropriate to relate the insulating phases to the $B=0$ metal-insulator transition \cite{coleridge97b}, where the disorder and Coulomb interaction terms are the largest energies involved, than to quantum Hall states where the cyclotron energy dominates.

Despite the strong similarities between the magnetic field induced insulating phases in p-SiGe and Si MOSFETs there are probably additional factors involved for p-SiGe heterostructures. The Zeeman splitting in p-SiGe is large, $g\mu_BB$ is of order $\frac{1}{2}\hbar\omega_c$ and is further increased by exchange enhancement \cite{ando82} so $g_{\text{eff}}\mu_BB$ may exceed $\hbar\omega_c$. The ordering of the Landau levels then changes and, in particular, the second and third (spin resolved) levels cross giving a ferromagnetically polarised system at $\nu=2$, as was experimentally shown in Ref.\cite{coleridge97a}.  If the magnetic field is further increased the exchange enhancement of the Zeeman splitting has saturated but the cyclotron energy continues to increase. There is then the possibility that the Landau levels may cross again while reverting back to their normal configuration. There is some indication \cite{coleridge97a} that the insulating phase $\nu=\frac{3}{2}$ only appears if this crossing occurs. Then the appearance of re-entrant insulating phases would depend on the magnitude of the exchange enhancement of the Zeeman splitting. The lower the density the lower the magnetic field required to produce a large exchange enhancement and the increased probability that Landau level crossings occur at values of $\nu \geq 2$.

As noted previously, a re-entrant insulating state at $\nu\approx\frac{3}{2}$ does not necessarily contradict the topology of the global phase diagram of Kivelson, Lee, and Zhang \cite{kivelson92} shown in Fig.~1~(a). This is not the case, however, for the other two insulating phases which appear between QH states belonging to upper Landau levels and therefore contradict the genuine global phase diagram.  Topologically, they are permitted by the phase diagram of Sheng and Weng \cite{sheng00} (Fig.~1~(c)). The main difference between this and the original diagram is that the upper extended states (lines separating QH states with different non-zero values of $\sigma_{xy}$) merge with the lowest extended state (line separating the insulating state from the QH states), instead of nesting.  (Experimentally, this merging was observed for the first time in Si MOSFETs by Shashkin {\it et al.} \cite{shashkin93}.)  As we have already mentioned, to accommodate multiple {\it field-induced} QH-insulator-QH transitions, the phase diagram must have the ``bumpy'' phase boundary shown schematically in Fig.~1~(d), again as in Si MOSFETs \cite{diorio90,pudalov93,shashkin93,kravchenko95}.  Sections of this phase diagram corresponding to the $\rho_{xx}(B)$ dependences of Fig.~2~(a) and (b) are depicted by the dashed and dotted lines respectively. 

In summary, we have observed, for the first time in p-SiGe heterostructures, multiple field-induced QH-insulator-QH transitions similar to those previously observed in dilute Si MOSFETs.  These transitions are therefore a universal property of dilute 2D systems.  Some of them are ``forbidden'' by the original global phase diagram of Kivelson, Lee, and Zhang \cite{kivelson92}.  They are topologically permitted by the numerical diagram of Sheng and Weng \cite{sheng00}, although their diagram also requires some modifications.  As in Si MOSFETs it appears that the spin of the carriers plays an important role in the appearance of these insulating phases.

M.~R.~S., M.~R., and S.~V.~K. are supported by NSF grants DMR-9803440 and DMR-9988283 and Sloan Foundation.
     
\end{multicols}


\begin{thebibliography}{10}
\bibitem{sampledetails} The wafer was CVD192 as described in Ref.\protect\cite{coleridge97b}.
\end{thebibliography}
\end{document}